# Microcavities Using Holey Fibers

S.M. Hendrickson[1,2], T.B. Pittman[1], and J.D. Franson[1]

[1]*Physics Department, University of Maryland, Baltimore County, Baltimore, MD 21250*
[2]*Electrical and Computer Engineering Department, Johns Hopkins University, Baltimore, MD 21218*

**Abstract:** Vacuum compatible microcavities consisting of microstructured holey fibers and separate end mirrors have been constructed and tested. These devices exhibit excellent transverse mode confinement and the ability to control the percentage of power guided outside of the fiber core. As a result, these devices may be a useful tool for enhancing the interaction between light and an atomic medium.

The increase in the field intensity that occurs within a resonant optical cavity has been used in a variety of applications to strengthen the interaction between light and matter [1]. Recently, various types of photonic crystal or holey fibers have been used to provide a high degree of transverse confinement, effectively increasing the strength of the field within these waveguides. Here we demonstrate a new type of microcavity that combines both types of confinement to enhance the field strength, with the added benefits of an arbitrarily small cavity length and an evanescent field strongly coupled to an atomic vapor. Microcavities of this kind may be useful for investigating nonlinear effects at low intensities and they may have applications in quantum logic gates. [2,3,4].

Other types of microcavities, such as microtoroids and microspheres [5], exhibit very high quality factors (Q values) and small mode volumes but most of the field is confined within the dielectric material. As a result, these devices may require laser trapping to position individual atoms at optimal positions outside the device in order to achieve a sufficiently strong interaction. While much success has been demonstrated with these devices in experiments with small numbers of atoms [6], the holey fiber cavity represents a possible improvement for experiments focused on observing effects between a small number of photons and an ensemble of atoms in a vapor, since small mode volumes can be achieved with more power outside of the dielectric.

The experimental arrangement used to investigate the properties of holey fiber cavities of this kind is shown in Fig. 1. The resonator consisted of a short section of holey fiber (20.5 mm) with partially-reflective mirrors pressed against the cleaved ends of the fiber. The holey fiber used in these experiments was Crystal Fibre NL-15-670-02, which guides light by total internal reflection in a 1.5 $\mu m$ silica core surrounded by air holes as shown in Fig. 2. This type of fiber strongly confines guided modes as a result of the large index difference between the silica core and the air cladding in the microstructured region. The microcavities were characterized at wavelengths near 778 nm to coincide with planned future experiments on two-photon absorption using rubidium vapor.

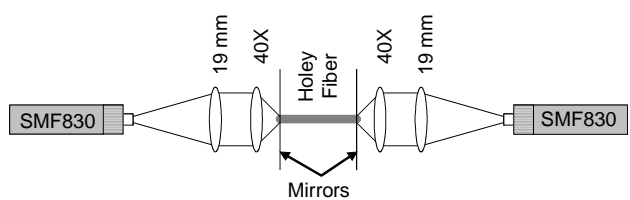

Fig. 1. Basic experimental arrangement used to study a microcavity formed by placing a short section of holey fiber between two partially-reflective mirrors. Two conventional optical fibers were coupled to the cavity using a series of lenses and micropositioners. Details can be found in the text.

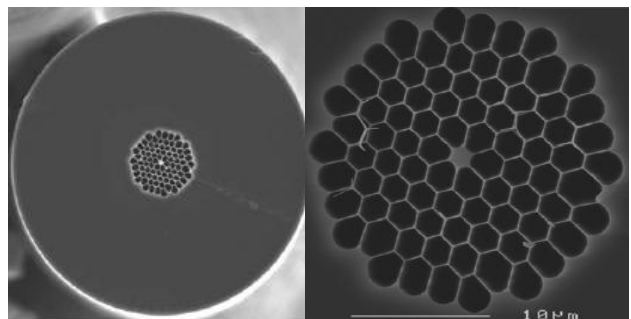

Fig. 2. A scanning electron microscope (SEM) image of the end-face of the holey fiber used in these experiments (courtesy of Crystal Fibre). An on-site SEM was used to verify the integrity of the microstructured region after cleaving.

The modal properties of the fiber used in these experiments are similar to those of a small-diameter silica rod suspended in air or vacuum [7]. Since there is no cladding in fibers of this kind, the evanescent field overlaps with any material that lies outside of the 1.5 $\mu m$ core. Of particular interest is the ability to insert atomic alkali vapors into these air-hole regions [8]. Interactions between guided photons and atomic vapors can be optimized by controlling the fraction of the field energy that propagates in the evanescent field. At 778 nm, this holey fiber supports multiple transverse modes with more energy located outside of the core for each higher order mode [9,10]. If operation in the fundamental mode is desired, a reduction of the diameter of the core will increase the power guided in the air or vacuum, as we will show. Alternatively, the field distribution can be controlled without modifying the fiber by an appropriate choice of wavelength relative to the core diameter.

The optical coupling setup shown in Fig. 1 was used to characterize the fiber cavity while allowing future placement of the device into a vacuum chamber. The output of a standard single-mode optical fiber (SMF830; Thorlabs P1-830A-FC-5) was collimated using an achromatic doublet lens (F =19 mm) and then focused onto the core of the holey fiber using a 40X microscope objective, which compensated for the different mode field diameters. In future applications, the windows of the vacuum chamber will be located in the collimated regions of the input and output beams in order to minimize any distortions.

An automated coupling system controlled by a computer was used to position the input and output fibers to maximize transmission through the holey fiber cavity. A three-axis positioner held each of the two SMF830 fiber launchers and each axis was controlled by an electrostrictive actuator capable of repeatable sub-micron accuracy. An automated algorithm was created to iteratively optimize the positioners while accounting for the hysteresis inherent to the actuators. The measurements reported here were all performed in air using a frequency-stabilized diode laser operating near 778 nm.

Both silvered and dielectric mirrors have been tested with reflectivities ranging from 90% to 99%. The mirrors were fabricated on microscope cover slips to allow some flexibility in the shape of the surface when pressed against the end of the fiber. The mirrors were gradually moved towards each end of the fiber using motorized translation stages while the output intensity of the system was monitored and re-optimized. After the mirrors contacted the ends of the fiber, the frequency was tuned to maximize cavity transmission and final adjustments were made to optimize the output signal.

The dielectric mirrors did not perform as well as the silvered mirrors for this application. We suspect that this is because the holey fiber has a high numerical aperture and light is emitted from the end of the fiber into a large angle, whereas the reflectivity of the dielectric mirrors is optimal for light of near normal incidence. This can be further understood from the fact that dielectric mirrors contain multiple boundaries that together create constructive interference in the reflected direction and destructive interference in the transmitted direction. As the light emerges from the fiber, it will diverge laterally and will not be refocused properly into the fiber. As a result, all of the data reported here was obtained using partially silvered mirrors.

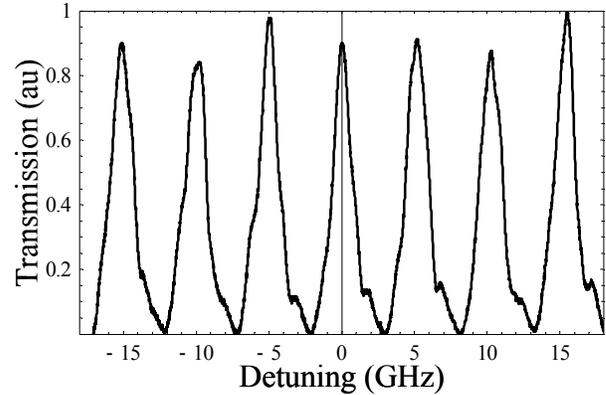

Fig. 3. Power transmitted through the holey fiber microcavity as a function of frequency detuning near 778 nm.

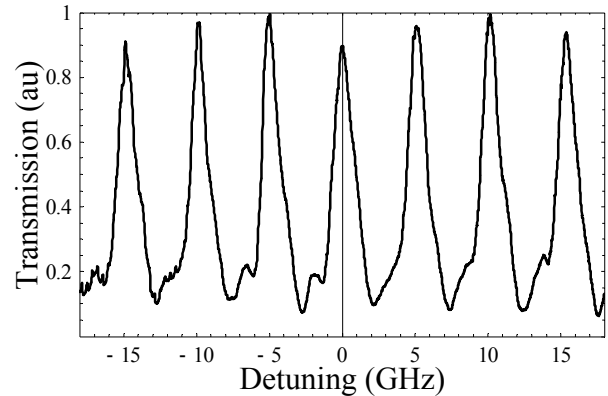

Fig. 4. Power transmitted through a single-mode fiber microcavity as a function of frequency detuning near 778 nm.

A plot of the power transmitted through a cavity consisting of a holey fiber and 90% silvered mirrors as a function of the laser wavelength is shown in Fig. 3. The power was measured using a Hamamatsu HC220-01 silicon photodiode. The wavelength was scanned at a rate of 1.25 GHz/s to allow the low-bandwidth detector to resolve all of the features of the transmission data. A scan of a single-mode fiber (Thorlabs SM800-5.6-125) cavity of the same length (20.5 mm) is shown for comparison in Fig. 4. The holey fiber scan shows that with proper coupling most of the power can be coupled into the fundamental mode of the fiber. It should be noted, however, that even with the coupling system described above some plots exhibited multiple-peak structures indicative of the excitation of multiple transverse modes within the fiber. Most of the planned applications of this cavity would require single-mode operation

The quality factors of these resonators ranged from $10^5$ to $10^6$ and were predominantly limited by losses at the interface between the mirrors and the fiber cavity. In addition to the transmission through the mirrors, loss can occur if the mirror is not aligned perfectly parallel to the end of the fiber or if the fiber cleave is not perfectly flat. This significant source of intra-cavity loss made it impractical to obtain meaningful data for mirrors with reflectivities higher than 90% since the fiber-mirror interface losses increase with each round-trip within the cavity. The effects of mirror misalignment are illustrated in Fig. 5. This source of loss could be mitigated by increasing the length of the microcavity, but increasing the cavity volume would reduce the effectiveness of this device for cavity quantum-electrodynamics (QED) experiments.

Other groups have demonstrated fiber cavities with conventional fiber and separate end mirrors with very low losses [11], but the techniques used are not compatible with holey fibers or high-vacuum conditions. Specifically, the ends of conventional fiber were polished and then coated with index-matching gel. Our application precludes these methods because polishing would likely damage the fragile inner structure of the holey fiber while index matching gel would clog the air-holes and outgas in high-vacuum conditions. Optical cavities have also been formed with holey fibers using fiber Bragg gratings [12]. The use of these reflectors would reduce intra-cavity losses as compared to external mirrors but mode volumes would be limited by the required length of the inscribed gratings.

The measured free spectral ranges (FSR) of the holey fiber cavity shown in Fig. 3 and the single mode fiber cavity shown in Fig. 4 were 5.15 GHz and 5.00 GHz, respectively. To calculate the expected values of these parameters, the following expressions for the effective index of the fundamental mode and the FSR were used [13,14]:

$$n_{eff} = \frac{\beta}{k_0} \quad FSR = \frac{c_0}{2n_{eff}L} = \frac{c_0 k_0}{2\beta L} = \frac{c_0 \pi}{\beta \lambda_0 L}$$

Here $k_0$ is the wave number in a vacuum, $\beta$ is the propagation constant of the fundamental mode of each fiber and L is the length of the cavity. For SMF830, the index difference between the core and the cladding is small and $\beta \approx n_{core}k_0$. For holey fibers the index difference is sufficiently large that the propagation constant must be found numerically by solving the eigenvalue equation that results from applying appropriate boundary conditions to Maxwell's equations [9]. The calculated values are shown in Table 1, which show reasonable agreement with the measured data:

Table 1. Calculated fiber cavity parameters

| | $\beta$ (µm)$^{-1}$ | $n_{eff}$ | FSR (GHz) |
|---|---|---|---|
| SMF830 | 11.74 | 1.45 | 5.03 |
| Holey Fiber | 11.38 | 1.41 | 5.19 |

The length of fiber used in these cavities was chosen to be compatible with fiber holders measuring 19 mm in width. The length of the cavities could be reduced to much smaller dimensions by cleaving shorter sections of fiber, or perhaps by temporarily imbedding the fiber in a plastic matrix and polishing it to a smaller length. This would substantially reduce the mode volume of the cavity.

For experiments with single photons and atomic vapors it is essential to maximize the overlap of the field with the location of the atoms, while at the same time reducing the mode volume. A plot of the theoretical mode volume, defined as the spatial integral over the field intensity normalized to unity at the field maximum [15], for a cavity of length 1 mm is shown in Fig. 6a for the fundamental mode of holey fiber. The minimum near 440 nm represents the diameter for which the light is most strongly confined at a wavelength of 778 nm. The value of the mode volume at this minimum scales linearly with the length of the cavity. The plot in Fig. 6b shows that this type of fiber has 49% of its power in the air or vacuum at the diameter of peak mode compression, making this an excellent tool for experiments studying the interaction of light with multiple atoms. These diameters could be achieved by heating and tapering the holey fiber [16], for example.

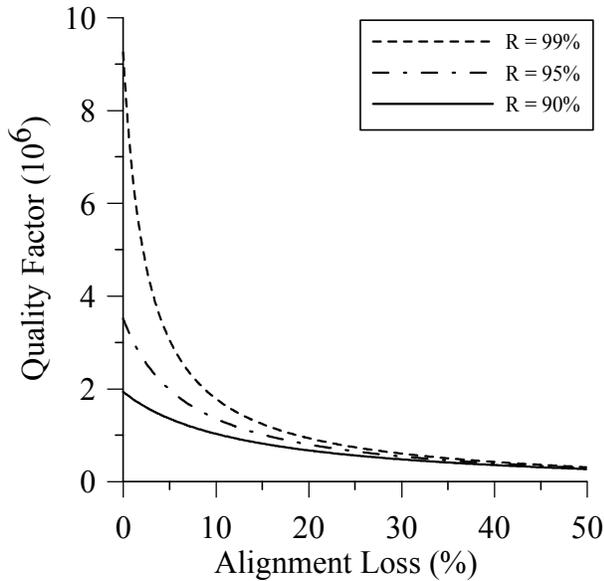

Fig 5. A plot showing how the alignment loss at each fiber-mirror reflection affects the quality factor of a 20 mm long section of holey fiber for three mirror reflectivities.

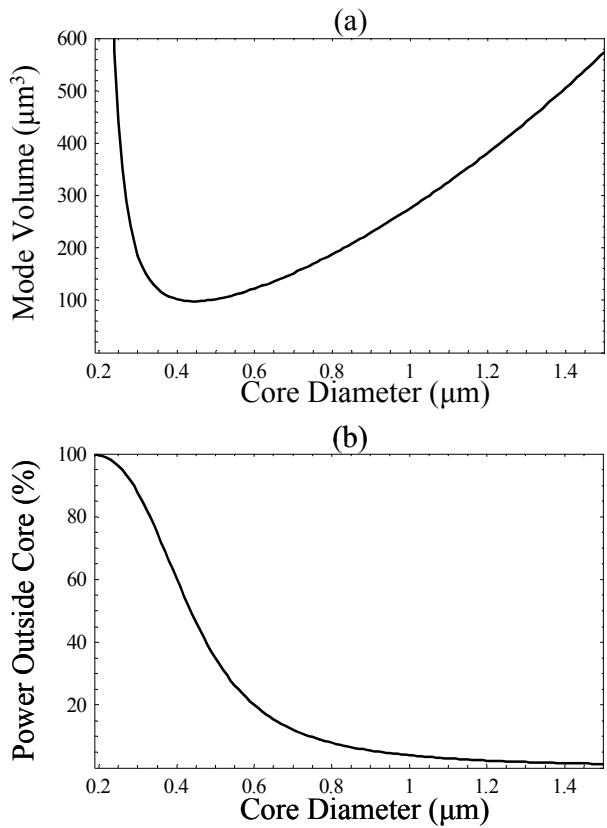

Fig 6. The transverse field distribution can be optimized by choosing the core diameter. (a) The mode volume and (b) the percentage of power guided in air for a cavity of length 1 mm at a wavelength of 778 nm.

In summary, we have built and tested a microcavity consisting of a short section of holey fiber with separate end mirrors that is suitable for high-vacuum applications. This cavity may be a useful tool for investigations of nonlinear optical effects at low intensities, and for QED and quantum logic experiments if the cavity losses can be reduced.

We would like to acknowledge helpful discussions with Jin Kang regarding these experiments. This work was supported by DTO-funded U.S. Army Research Office grant No. W911NF-05-1-0397.